\documentclass[11pt,twoside]{article}
\usepackage[hmarginratio=1:1,top=32mm,columnsep=20pt]{geometry}
\usepackage{multicol}
\usepackage[footnotesize,labelfont=bf,up,textfont=it,up]{caption}
\usepackage{float}
\usepackage{booktabs}
\usepackage{paralist}
\usepackage{amsmath}
\usepackage{wasysym}
\usepackage{graphicx}
\usepackage{hyperref}

\usepackage{abstract}

\usepackage{titlesec}
\renewcommand\thesection{\Roman{section}}
\renewcommand\thesubsection{\Roman{subsection}}
\titleformat{\section}[block]{\large\scshape\centering}{\thesection.}{1em}{}
\titleformat{\subsection}[block]{\large}{\thesubsection.}{1em}{}

\usepackage{fancyhdr}
\title{\textsc{Non-Cooperativity in Bayesian Social Learning}}
\author{Stan Palasek\\Princeton University}
\date{July 2014}
\begin{document}

\maketitle

\thispagestyle{fancy}

\begin{abstract}
\vspace{-.1in}
We describe a Bayesian model for social learning of a random variable in which agents might observe each other over a directed network. The outcomes produced are compared to those from a model in which observations occur randomly over a complete graph. In both cases we observe a nontrivial level of observation which maximizes learning, though individuals have strong incentive to defect from the societal optimum. The implications of such competition over information commons are discussed.
\end{abstract}

\begin{multicols}{2}

\section{Introduction}

Trends and fads, though at times ill-informed, are ubiquitous features of human societies. The mathematical examination of the origins of these trends as cascades of questionable wisdom was perhaps pioneered in 1992 by Banerjee~\cite{banerjee} and independently Bikhchandani \textit{et al.}~\cite{bikhchandani} Although it is rational for agents to observe the actions of others and to update their own beliefs accordingly, Banerjee notes the dilemma that ``the very act of trying to use the information contained in the decisions made by others makes each person's decision less responsive to her own information and hence less informative to others.'' Vives similarly showed that when agents' observations are governed by a benevolent organizer, social learning is more effective than when each individual is acting strictly in his or her own interest. The observed inefficiency is a manifestation of an ``underinvestment in production of public information,'' which limits exploration of unpopular decisions and propagates herding.~\cite{vives} These results may go a long way in explaining the emergence of societal trends which may seem irrational to those with access to a different stream of information. Indeed, theories concerning social learning have found a number of applications including finance,~\cite{welch,chamley} policy,~\cite{dobbin} and the adoption of new ideas and technologies in the developing world.~\cite{conley,munshi,banerjee13}

\section{Social learning model}

We consider a system similar to that in Salge \& Polani~\cite{salge}. Here, there are $n$ slots, of which one hides a reward which is being sought by $n$ agents. At each time step, with probability $p\ll1$ the reward is randomly relocated. A random agent $a$ is then selected uniformly and independently who observes the slot $s$ that it believes most likely to contain the reward. All $m$ agents perform a na\"ive Bayes update: the $a$th agent's based on the result of its observation of $s$, and the other agents' based on the fact that $a$ elected to observe $s$. This update takes into account the possibility that the target has moved, removing the risk that agents get ``stuck'' having unsuccessfully checked all $n$ loci.

As discussed in~\cite{salge}, an agent's actions are chosen in consideration of its previous observations and therefore contain potentially useful information. For instance, in a simple model with no Bayesian updating and $p=0$, it can be shown that the conditional probability of the reward being at $s$ given that an agent chooses to observe $s$ is twice the conditional probability given that an agent choses to observe a slot other than $s$. We might expect that introducing social updating only injects more useful information into agents' choices.

Described so far has been a model for social learning where the underlying social network is the complete graph of order $m$. More generally, we can consider the same process but with agents constrained to perform Bayesian updates only for those with whom they are connected on a graph $G$. Note that at times we will allow $G$ to be directed; ie.\ for some (perhaps artificial) reason, an agent $i$ employs information from $j$ but not \textit{vice versa}.

We make several key changes to Salge \& Polani's model which we believe make this formulation more realistic. First, here we do \textit{not} reset the beliefs of an agent once it discovers the location of the reward. We admit that occasionally injecting maximum entropy makes sense when $p=0$ since the game would otherwise be ``solved'' and quickly reach a steady state. However, the fact that the location of the reward changes simultaneously for the entire ensemble suggests that all the agents are within a single environment and it would therefore be unnatural for some to abruptly forget information or be replaced.

Second, here we use a social network to limit information consumption (as in \cite{bala,gale,acemoglu}) rather than throwing out observations with a uniform probability $1-p_o$ (reminiscent of \cite{smith}). Although the latter model is appealing in the sense that it provides a linear continuum between a non-social situation ($p_o=0$, or all isolated nodes) and complete observability ($p_o=1$, or $G$ is complete), it is far more natural for agents to form social networks and learn from those with whom they are connected. Indeed, one might expect the uniformly random information selection to be analogous to learning on an Erd\H{o}s-R\'{e}nyi random graph since they must both exhibit low diameter, clustering, and a binomial degree distribution. The authors suggested the random ignorance model as something that social agents might form strategically rather than something that is inherent within their decision-making processes, the plausibility of which we will investigate later.

Finally, here we order the agent observations uniformly and independently rather than in a predetermined sequence. This change was necessitated by the present use of the network model in lieu of random observation. The latter formulation is symmetric because the choice of information disposal is i.i.d. The arbitrary designation of an order is therefore of little concern. In the current model, however, the underlying graph is generally asymmetric so an arbitrary ordering of the agents may introduce harmful bias. Random ordering avoids this complication.

\section{Graph Connectivity Results}

A major result of \cite{salge} is that maximal learning occurs at some intermediate $0<p_o<1$. We will illustrate an analogous fact for our network-based model with the modifications detailed above. In particular, we execute the described process for graphs chosen uniformly at random from the set of graphs with $n$ vertices and $i$ edges, for each $i=1,2,3,\ldots,\binom{n}{2}$.
\begin{figure}[H]
\includegraphics[scale=.252]{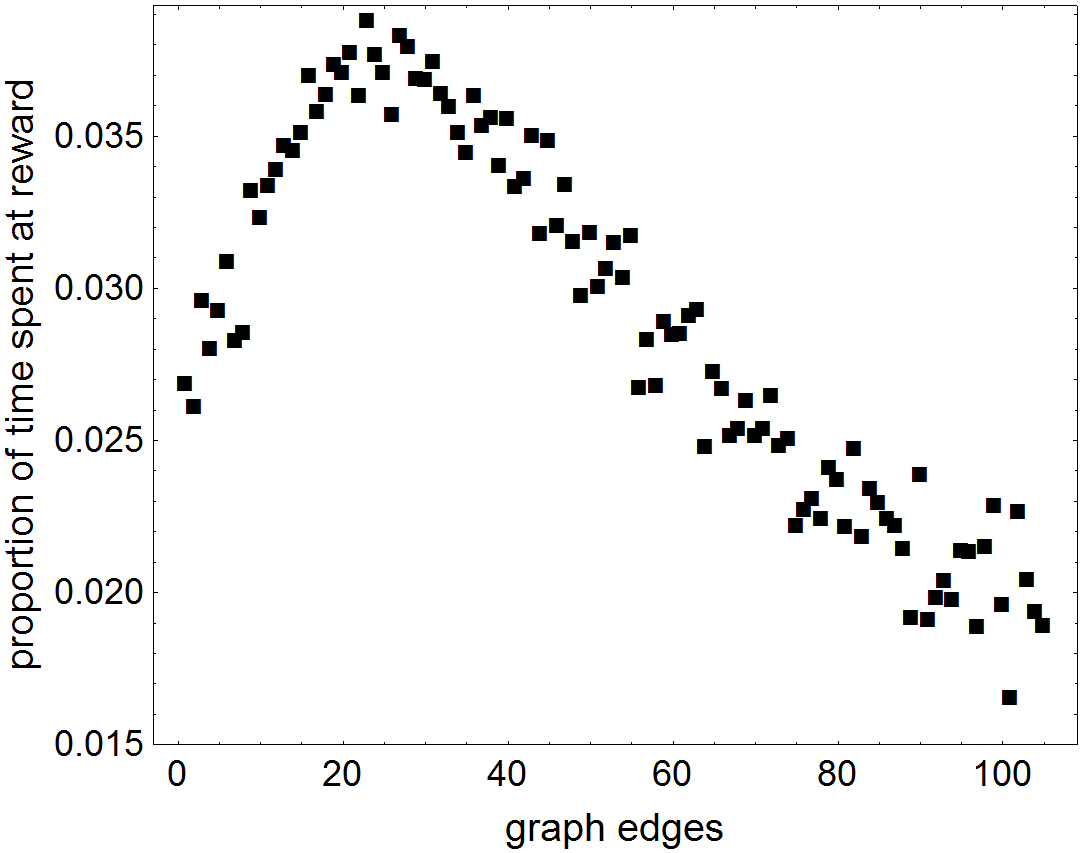}
\caption{Simulation results for $n=15$, $p=0.01$, and a Bayes' factor of 2 between positive and negative observations. Depicted are the results of 50 trials each a mean of 1000 iterations at each edge count from 1 to 100.}\label{edgecount}
\end{figure}

\begin{figure}[H]
\includegraphics[scale=.303]{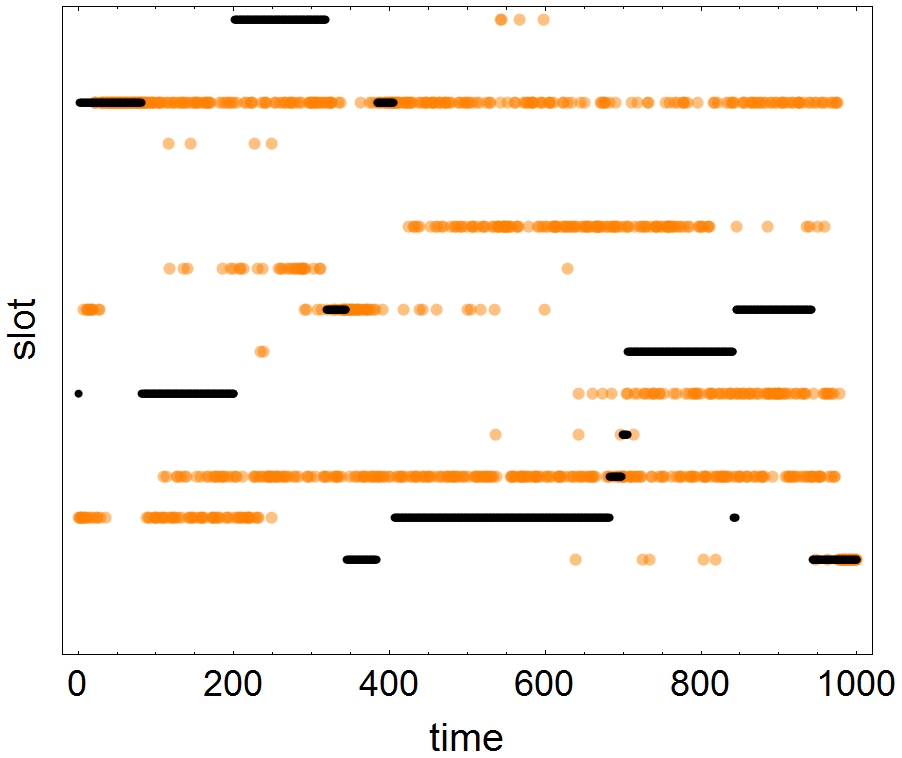}
\caption{The location of the treasure (black) and the slot chosen at a given time (orange) for the same parameter set as Figure~\ref{edgecount}. Representative data for a society over a complete graph are shown, and misinformed herding behavior is evident.}\label{herding}
\end{figure}

The results are shown in Figure~\ref{edgecount}. We see, as in \cite{salge}, that there indeed exists an optimal quantity of information to be collected by social agents, as learning is maximized when the graph has approximately 25 edges, or when the mean degree is $3.3$. For very disconnected networks, agents have little access to the information contained in others' actions which, as we argued, has a Bayes' factor of at least two. Conversely, when the network has most of its possible edges present, we observe that herding behavior to incorrect slots is prevalent. This is illustrated in Figure \ref{herding} which depicts the simulation over a complete network during which the agents spend most of their time at only a handful of spots which rarely contain the reward.

On these grounds alone, \cite{salge} suggests that rational agents might voluntarily ignore information to achieve $p_o<1$ or, in this setting, a non-complete graph. In the next sections we will evaluate whether agents truly have any incentive to do so, or whether defecting from an optimal configuration would suit one's individual interests at the expense of those of the group.

\section{Stability over Complete $G$}\label{completesec}

The notion that agents freely decide whose information to take into account and to what extent suggests that by default they have at their disposal more information than they would optimally use. In this section we will therefore take the complete graph as a starting point of our investigation of stability and consider deviations thence. Should we see that from here no one has incentive to discard information, then we must conclude that the complete graph constitutes a stable equilibrium. However, the prevalence of harmful herding as demonstrated in the last section would lead one to expect that agents are eager to defect from complete networks so as to minimize their susceptibility to cascades of false information. We test this by modifying a complete graph such that a test node, originally of degree $m-1$, instead has degree $i$ for $i=0,1,2,\ldots,m-1$. Since the original graph was complete and therefore symmetric, there is a unique unlabeled graph for each $i$. This will be an instance, as we hinted at earlier, in which the graph is directed (except when $i=14$) since an agent's decision to disregard data from its neighbor has no bearing on its neighbor's reciprocal observation.

\begin{figure}[H]
\includegraphics[scale=.255]{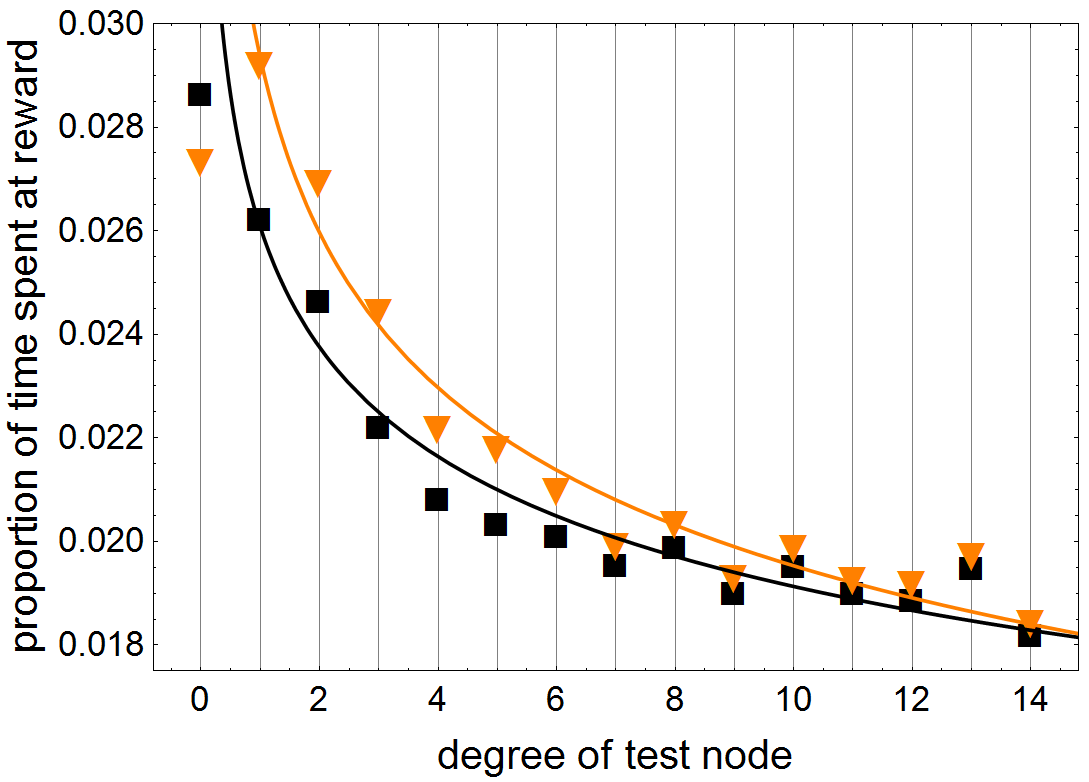}
\caption{Simulation results for every possible single-agent deviation from a complete graph. Depicted are time series for both the defecting agent (orange) and the remaining $m-1$ collectively (black). The parameter set is the same as that from Figure~\ref{edgecount}. Both series for $t\geq1$ follow power laws with $R^2=0.999$ with forms $P=0.0293\times d^{-0.177}$ (individual) and $P=0.0261\times d^{-0.134}$ (society). We do not show data for the individual non-defecting agents for they vary in proportion with the aggregate data.}\label{complete}
\end{figure}

The result of the simulation is shown in Figure~\ref{complete}. As expected, single agents are generally rewarded for severing ties from a complete network. The only robust exception is in the transition from maintaining one connection to becoming isolated, in which the defecting node takes a slight loss in performance by disconnecting itself completely. Furthermore, we see that the improvement in the defector's performance is mirrored in the welfare gain experienced by the rest of society. We must then ask, given that deviations from the complete network are incentivised, to what density the network must be cleared for no defection to be motivated, should such a stable network exist at all.
\section{Stability over Sparse $G$}

\begin{figure}[H]
\includegraphics[scale=.235]{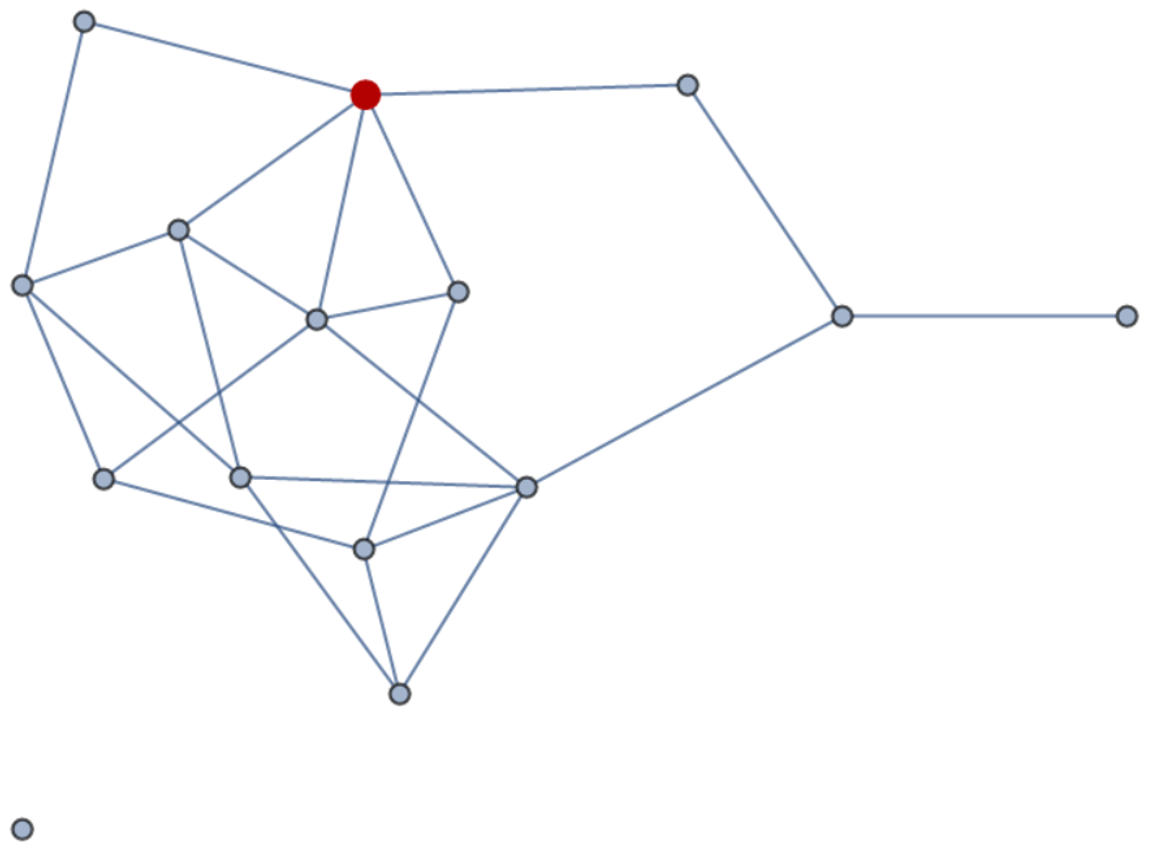}
\caption{$G_0$, the highest-performing graph with edge count near 25 among the 100,000 sampled graphs depicted in Figure~\ref{edgecount}. The ``test vertex'' is highlighted in red.}\label{g}
\end{figure}

Next, rather than beginning with a complete graph, we will instead consider the the highest performance graph with edge count near the extremum of Figure~\ref{edgecount}, chosen out of the 100,000 random 15-order graphs examined. This graph, call it $G_0$, is depicted in Figure~\ref{g} and it may be near the equilibrium, should one exist.
For the purpose of illustration we will arbitrarily designate the node highlighted in red as the test vertex which will defect from its current situation in $G_0$ during our experiments. In particular, we will repeat the simulation from the previous section upon this graph $G_0$ by varying the order of the text vertex. This procedure differs in that now inasmuch as the starting graph is asymmetric, there is \textit{not} a unique graph with the required properties. We will therefore select a uniformly random degree inclusively between 0 and $m$ and then choose the network randomly from the class of graphs in which the test vertex has the selected degree and connections between non-test vertices are unchanged.

\begin{figure}[H]
\includegraphics[scale=.3]{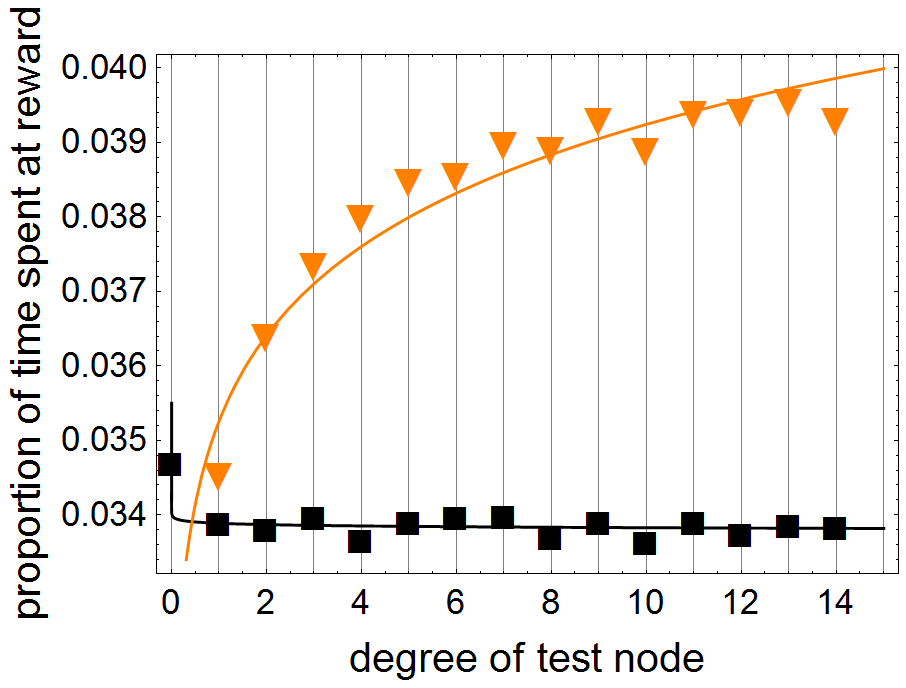}
\caption{Learning performance over optimal graph $G_0$ with the test model's connections systematically modified. 50,000 trials were executed with the same parameter set used in Figure~\ref{edgecount}. Depicted are both the performance of the test node (orange) and the rest of the network (black). Both time series are fit to power laws with $R^2=0.9999$ with forms $P=0.0353\times d^{0.0466}$ (individual) and $P=0.0339\times d^{-0.00762}$ (society).}\label{optimal}
\end{figure}

\begin{figure}[H]
\includegraphics[scale=.25]{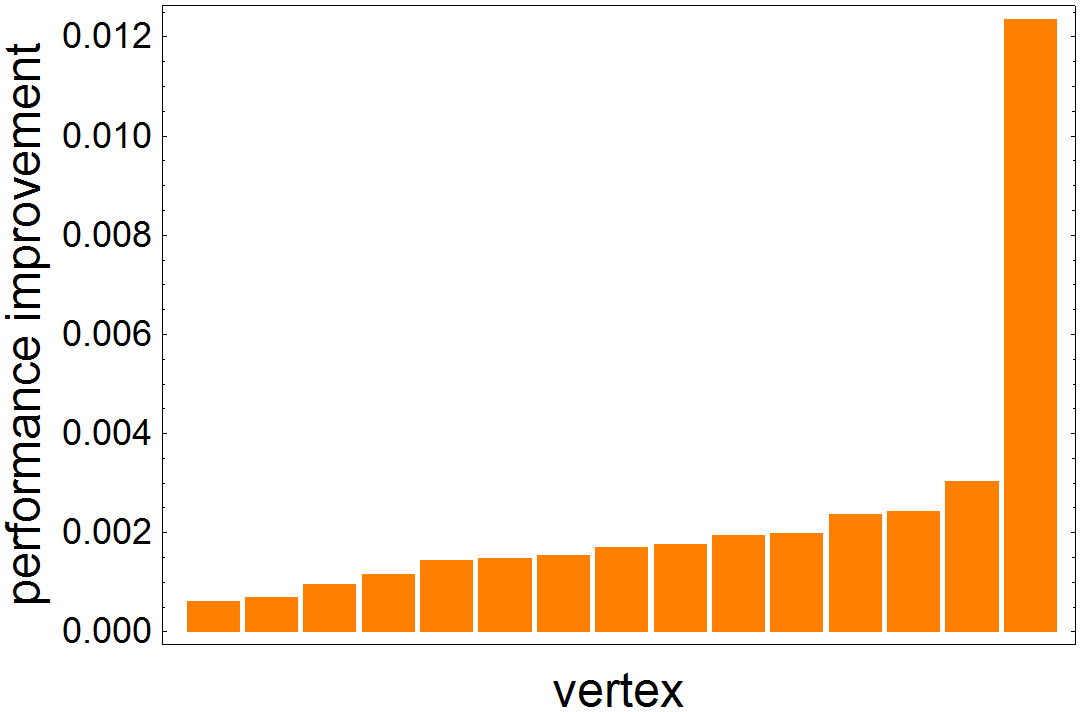}
\caption{Mean performance improvements for each node of $G_0$ upon independently defecting to complete observation. The same parameter set from Figure~\ref{edgecount} is employed with 1000 trials for each node. The outlying datum is that of the isolated node seen in Figure \ref{g}.}\label{diffs}
\end{figure}

Figure~\ref{optimal} depicts the results, which show that the test node can expect a robust increase in its average performance by observing more agents. Meanwhile, as long as the test agent is connected to \textit{someone}, its behavior ceases to have a measurable effect on the success of the others. Because the test agent, already occupying one of the better-connected nodes in the network, is motivated to expand his observation of the other agents, we conclude that this network does not present a stable environment when individuals are free to adjust their information consumption. Figure~\ref{diffs} confirms that this observation is a general one and that all nodes--not just the test node of Figure~\ref{g}--are motivated to defect to complete observation.

\section{Stability with Random Observation}

One might come to believe that the instability we have identified when individuals are free to choose which information to consider is a feature of our graph model and is not present in the random observation model of \cite{salge}. Here we will show that the phenomenon is present regardless.

\begin{figure}[H]
\includegraphics[scale=.253]{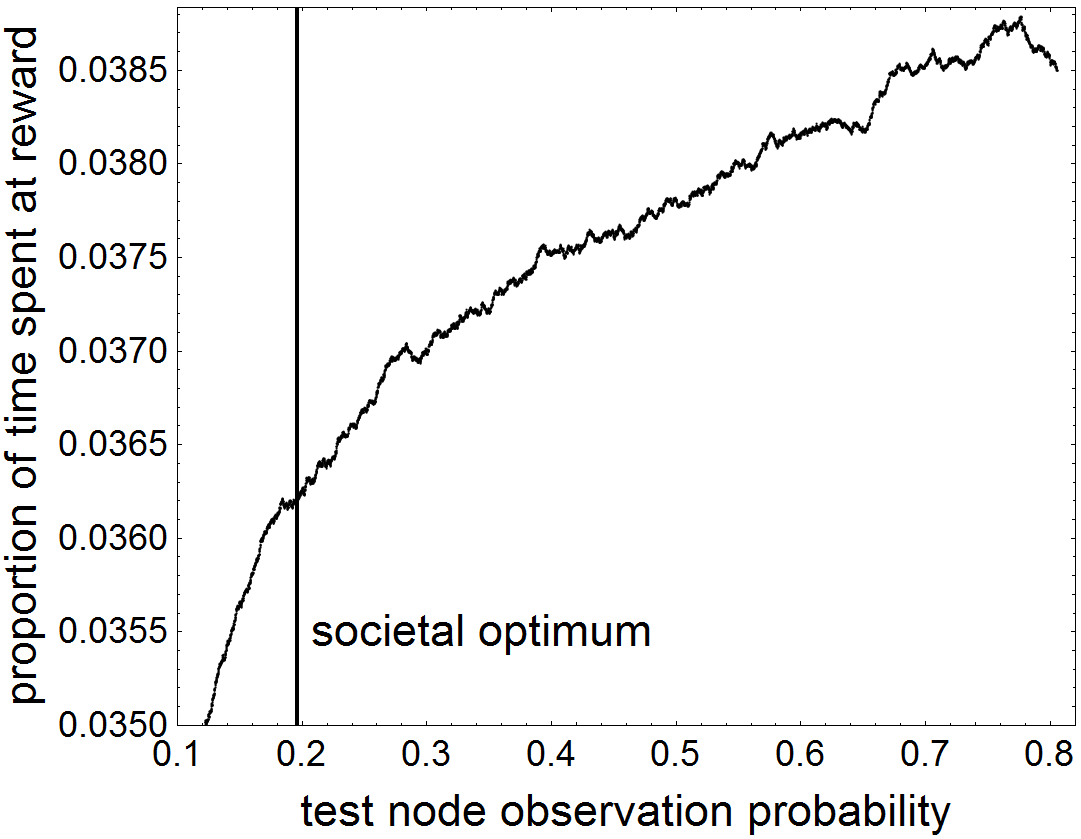}
\caption{Learning performance for test node as it defects from the societal ideal $p_o=0.196$. 10,000 trials were executed, each time selecting a random observation probability for the test node. Data are subjected to exponential moving averaging with a smoothing factor of $0.0005$.}\label{obsprob}
\end{figure}

Now instead of each agent forming its beliefs according to its connections in a directed graph, it will use information from a given agent at a given time with some probability $p_o$. As noted before, varying $p_o$ between 0 and 1 interpolates between the null and complete graphs. In order to understand the incentives of agents to deviate from the $p_o$ that is ideal for society as a whole, we first set all agents' observation probabilities to $p_0=0.196$, the optimal probability identified among 10,000 trials with the same parameter set that we have used throughout. We then allow the observation probability of a single test node to vary.\footnote{The choice of the test node does not matter because here there is no underlying graph to introduce asymmetry.} The results are depicted in Figure~\ref{obsprob}. We conclude that the test agent, and therefore every agent by symmetry of the model, has incentive to increase its observation probability far past the optimum for society. Therefore, despite $p_o=0.196$ being ideal for society as a whole, it will not be adhered to by rational agents who can freely choose their observation rates.

\section{Game Theory}

Up to this point we have been employing language as though we are examining a competitive game without being explicit about its components. It will become apparent as we do so how naturally our formulation can be viewed as a normal-form game.

Let the game have $m$ players corresponding to the society's agents. A pure strategy profile is any directed graph $G$ over $m$ labeled nodes in which the $i$th agent has control over the $i$th row of $G$'s adjacency matrix. The payoffs are defined as the average performance for each agent when the described simulation is carried out over $G$.\footnote{It is reasonable to expect that agents act upon the expectation of their utility.~\cite{bernoulli} For a treatment that does not neglect the stochastic nature of learning performance, one might consider games with ``fuzzy'' payoffs; see, for instance, \cite{clemente}} We assume that the agents' information on the simulation's rules, including the value of $p$, is complete.\footnote{One need not make this assumption. See Dekel \textit{et al.}~\cite{dekel} who examine learning without it.} In terms of this game, we can summarize the results of the previous two sections by noting that neither the complete graph nor $G_0$ is a Nash equilibrium since someone, and in fact everyone, gains a profit on average by defecting. Furthermore, the complete graph is Pareto inefficient by Figure \ref{complete}. We might wish to find a stable outcome, ie.\ a Nash equilibrium, but as we are considering only pure strategies, Nash's theorem does not guarantee their existence.~\cite{nash} To find equilibria exhaustively is intractable for the number of unlabeled undirected graphs when $m=15$ is on the order of $10^{19}$.~\cite{oeis} When directed edges are introduced the space becomes larger still by orders of magnitude.

\begin{figure}[H]
\includegraphics[scale=.378]{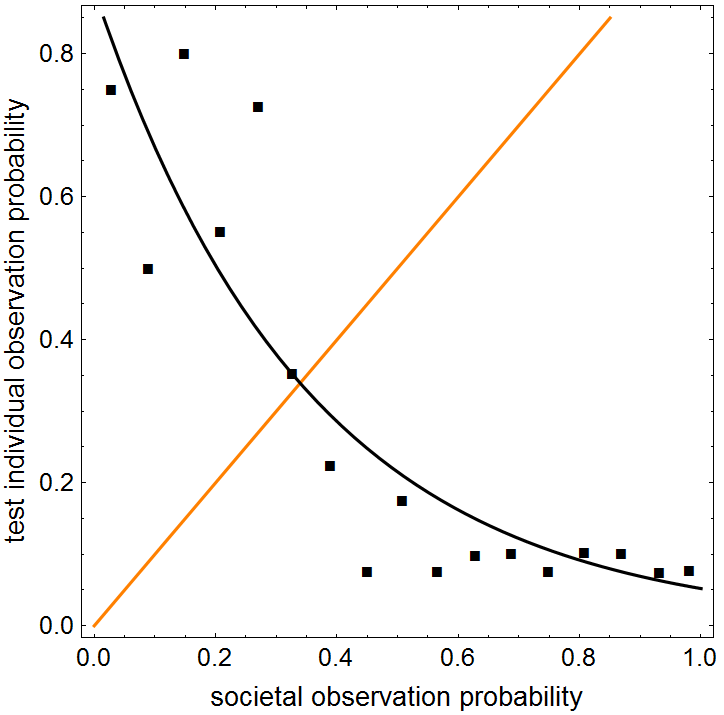}
\caption{The test node's best-response observation probability is depicted at various societal observation probabilities. The data represent 25,000 random samples of values for $p_s$ and $p_t$. The best response function is fitted to an exponential with $p_t^*=0.89\times\exp(-2.84\times p_s)$. \eqref{max} is satisfied when the best-response series (black) coincides with the locus $p_s=p_t$ (orange).}\label{eq}
\end{figure}

We may still, however, computationally identify an equilibrium within the observation probability model. Here, although the strategy space is continuous, it is one-dimensional and thus easy to search. We proceed as follows. Let $p_t$ be the observation probability used by the test individual, and $p_s$ the probability for the rest of society. Let the expected performance of the test individual be given by $U(p_s,p_t)$. In equilibrium, no individual has incentive to deviate from $p_s$. Therefore, we seek a solution $(p_s,p_t)\in[0,1]^2$ to
\begin{align}\label{max}
p_s=\operatorname*{arg\,max}_{p_t}U(p_s,p_t).
\end{align}
Clearly when $p_o=0$, there is no widespread risk of herding so the test agent's best-response is a $p_t$ near 1. Furthermore, as we showed in Section~\ref{completesec}, when $p_o=1$ the best-response is a near-zero level of observation. These facts combined with the continuity of the right-hand side of \eqref{max} and the intermediate value theorem would imply that there exists a solution to the equation. However, due to the prevalence of phase behavior as the degree of connectivity of societies varies,~\cite{erdos} we cannot be certain as to how the function behaves in practice. With this in mind we evaluate the function \textit{in silico}. The left- and right-hand sides of \eqref{max} are shown respectively in orange and black in Figure \ref{eq}.

\section{Discussion}

We, as in \cite{salge}, determine that maximal social learning occurs when not all information is taken into account by all agents. The authors go on to detail the reasonable corollary that rational agents will voluntarily discard much of their information--between $70$ and $80\%$ according to both our analyses. Doing so gives individuals' own observations greater weight in their Bayesian beliefs, thus globally limiting the impetus for herding which, as seen in Figure~\ref{herding}, is often to the wrong location. Our objection to this reasoning is as follows. A single agent, by defecting and letting the extent of its observation be universal (or nearly so), has little effect on the prevalence of herding, as it is necessarily a population-wide phenomenon. However, its defection yields in its favor a significant improvement in its ability to take advantage of emerging trends. Since the propensity for false information cascades has hardly increased, the defecting agent enjoys greater performance. Such an argument applies to any and therefore every member of the society. Consequently, we cannot expect rational individuals that have the ability to increase the extent of their observation to adhere to the welfare-maximizing level of information consumption.

This dilemma is highly reminiscent of Hardin's tragedy of the commons.~\cite{hardin} There is a common resource in our system: information uncontaminated by negative information cascades. Acting in its own self-interest, each agent is strongly motivated to maximally consume this information (see Figures \ref{optimal}, \ref{diffs}, and \ref{obsprob}). However, as the common pool of information becomes over-consumed, small random flocks to potentially incorrect loci are accentuated and herding behavior emerges; thus the resource is of far lesser use.

As Hardin famously pointed out, there is no technical solution to such problems of overconsumption. Fortunately, nature seems to impose a limit herself, for human societies are typically arranged in non-complete social networks. This raises the question of whether the properties exhibited by graphs of actual human communities are particularly conducive to such learning. One might also wonder, if such learning processes played any role in the adoption of technologies by early man which would have in turn correlated with survival probabilities, why defectors did not invade populations of mutual cooperators. For this we refer to the extensive literature on the evolution of cooperation in which these same questions are addressed in a general setting.~\cite{axelrod,nowak,ohtsuki}

\bibliographystyle{unsrt}

\end{multicols}

\end{document}